\documentstyle[aps,prd,subeqnarray]{revtex}
\newcommand{\beq}{\begin{equation}}
\newcommand{\eeq}{\end{equation}}
\newcommand{\beqn}{\begin{eqnarray}}
\newcommand{\eeqn}{\end{eqnarray}}
\newcommand{\nnb}{\nonumber}
\newcommand{\ppp}{\partial}
\newcommand{\V}{\nabla}
\newcommand{\EEE}{\epsilon}
\newcommand{\Lie}{\mbox{$\pounds$}}    
\newcommand{\ZZ}{\mbox{${\sf Z}{\hskip -0.3em}{\sf Z}$}}
\newcommand{\tri}{\triangle}

\newcommand{\brv}{\breve}
\newcommand{\bri}{\mbox{$\breve\imath^{\ssr +}$}}
\newcommand{\1}{{}^{^{(1)}}{\hskip-2.8pt}} 
\newcommand{\0}{{}^{^{(0)}}{\hskip-2.8pt}} 
\newcommand{\ov}{\overline}
\newcommand{\T}{\s{T}}
\newcommand{\s}[1]{{\scriptscriptstyle{#1}}}
\newcommand{\ssr}[1]{\scriptscriptstyle{\rm \, #1}}
\newcommand{\h}[1]{ {\hat{#1}} }
\newcommand{\hs}[1]{ \h{\hskip-0.3ex\ssr{#1}} }        
\newcommand{\bm}[1]{ \mbox{\boldmath{$#1$}} }          
\newcommand{\mmbox}[3]{\mbox{$\hspace{#1}\mbox{#2}\hspace{#3}$}} %
\newcommand{\up}[1]{{}^{^{(#1)}}{\hskip-3.0pt}}
\newcommand{\upp}[1]{{}^{\scriptscriptstyle{(#1)}}{\hskip-2.4pt}}
\newcommand{\straightup}[1]{ {}^{{}^{{}_{\ssr{#1}}}} {\hskip-3.3pt} }
\newcommand{\base}[2]{ {\scriptstyle (} #1_{\! \ssr{#2}} {\scriptstyle )} }
\begin{document}
\thispagestyle{empty}
{\baselineskip-4pt
   \font\yitp=cmmib10 scaled\magstep2
   \font\elevenmib=cmmib10 scaled\magstep1  \skewchar\elevenmib='177
 \leftline{\baselineskip20pt
 \hspace{12mm} 
 \vbox to0pt
    { {\yitp\hbox{Osaka \hspace{1.5mm} University} }
      {\large\sl\hbox{{Theoretical Astrophysics}} }\vss}}
%
 
 \rightline{\large\baselineskip14pt\rm\vbox to20pt
               {\hbox{OUTAP-85}
	        \hbox{DAMTP-1998-138}
		\hbox{UTAP-304}
                \hbox{RESCEU-52/98}
                \hbox{\today}
           \vss}
           }%
}
\vskip15mm
\begin{center}
{\large\bf Asymptotically Schwarzschild Spacetimes}
\end{center}

\begin{center}
{\large 
   Uchida Gen$^{1,3}$ and 
   Tetsuya Shiromizu$^{2,3,4}$\footnote{JSPS Postdoctoral Fellowship for Research Abroad}\\
\bigskip
\sl{ $^1$ Department of Earth and Space Science, Graduate School of Science,\\
          Osaka University, Toyonaka 560-0043, Japan,\\
                      \vskip3.5mm
     $^2$ DAMTP, University of Cambridge \\ 
          Silver Street, Cambridge CB3 9EW, UK, \\
                      \vskip3.5mm
     $^3$ Department of Physics, The University of Tokyo, Tokyo 113-0033, Japan,\\
                      and\\
     $^4$ Research Center for the Early Universe(RESCEU), \\ 
          The University of Tokyo, Tokyo 113-0033, Japan
  }
}
\end{center}

\begin{abstract} 
It is shown that if an asymptotically flat spacetime is
asymptotically stationary, in the sense that $\Lie_{\xi} g_{ab}$
vanishes at the rate $\sim t^{-3}$ for asymptotically timelike vector field $\xi^a$,
and the energy-momentum tensor vanishes at the rate $\sim t^{-4}$,
then the spacetime is an asymptotically Schwarzschild spacetime.
This gives a new aspect of the uniqueness theorem of a black
hole.

\end{abstract}
\vskip1cm

\section{Introduction}

There are many astrophysical phenomena that are best explained by black
holes, for example, active galactic nuclei and X-ray binaries. 
The analysis on the phenomena are done assuming that 
those black holes are described by the Kerr spacetimes.
This is because the uniqueness theorem
of a black hole guarantees that a spacetime which is stationary, 
vacuum and asymptotically flat is uniquely the Kerr spacetime,
and we believe that when gravitational collapse takes place and a black hole
is formed, the spacetime around it becomes vacuum and accordingly stationary.
(For the details on the uniqueness theorem of a black hole, see \cite{He}.)
However, one may argue that such a spacetime 
does not become {\em exactly} vacuum nor {\em exactly}
stationary: the spacetime becomes {\em asymptotically} vacuum, and
accordingly {\em asymptotically} stationary at a certain rate of the time.
In this context, a more adequate ``uniqueness theorem'' is the one 
that states an asymptotically
stationary, vacuum and flat spacetime is uniquely an asymptotically
Schwarzschild spacetime. This is what we show in this paper.
(One may find it peculiar that an asymptotically {\em stationary} spacetime
is an asymptotically Schwarzschild spacetime. However,
if we define {\em asymptotically} Schwarzschild spacetimes as a class of
spacetimes that asymptotically approach the exact Schwarzschild spacetime, the
spacetimes comprise a wide class of spacetimes, including the Kerr spacetime, 
which is stationary. See the end of sec.\ref{sec:stationary} for the details.)

To show the theorem, we use the notion of the asymptotic flat spacetime, 
first introduced by Ashtekar and Romano\cite{AR} at spacelike infinity
and succeedingly developed in our previous study\cite{GS}
at timelike infinity. We investigate the asymptotic behavior of the
gravitational field at the future timelike infinity,
because we would like to know whether the gravitational field approaches
asymptotically that of the Schwarzschild spacetime 
at the {\em late time} when the spacetime becomes asymptotically stationary.
The standard definition of the asymptotic flatness\cite{AH} 
is based on the conformal completion method,
which is used to obtain the Penrose diagram.
This method is useful in that spacelike and timelike infinities can be
simultaneously treated with null infinity, and thus that investigation concerning
the global causal structure can be done.
However, the method compresses the spacelike and timelike infinities,
which possess rich 3-manifold information, down to points, and
this compression results in 
a complicated differential structure at these points and makes it
difficult to obtain the comprehensive picture of the behavior of 
gravitational fields in general asymptotically flat spacetimes.
In contrast, the completion method introduced by \cite{AR} for
definining the asymptotic flatness at spacelike infinity
leaves the infinity as a 3-manifold.
As a result, the complicated differential structure in the former
treatment can be avoided.
Subsequently, in our previous study\cite{GS}, we applied the method to
timelike infinity and clarified 
that the method leads to a definite picture of hierarchy 
in the asymptotic behavior of the gravitational field and the symmetry,
and that it is suitable to discuss such a notion as an ``asymptotically
Schwarzschild spacetime''.
(Perng also investigates hierarchy in the asymptotic gravitational field
at spacelike infinity\cite{Pe}, 
although it does not directly correspond to the hierarchy discussed in \cite{GS}.)
In this paper, we further investigate the hierarchy and prove the
theorem stating that an  asymptotically
stationary, vacuum and flat spacetime is uniquely an asymptotically
Schwarzschild spacetime.

The plan of this paper is the following.
Sec.\ref{sec:pre} is devoted to preliminaries.
The definition of the asymptotic flatness at timelike infinity
is recalled,
and some of its useful consequences are summarized for the subsequent discussion.
In sec.\ref{sec:first}, the first order asymptotic structure is
explored, that is, the one order higher structure than the basic
asymptotic structure that all the asymptotic flat spacetimes possess.
Then, in sec.\ref{sec:stationary}, we introduce the notion of asymptotic
stationarity and investigate how the condition that an asymptotic flat
spacetime be asymptotically stationary, constrain the asymptotic
structure. The investigation leads to the proof of the main theorem.

Throughout the paper, we follow the notation of Wald\cite{Wa}. 

\section{preliminaries}\label{sec:pre}

In this section, we recall the definition of asymptotic flatness at
timelike infinity of \cite{GS} 
that will be used in the main proof and fix the notation.

\medskip

\noindent
{\bf DEFINITION:} \enskip A physical spacetime ($\h{\cal M}$, $\h g_{ab}$) is
said to possess an {\em asymptote at future timelike infinity $\bri$ to order n}
({\sc ati}-$n$) for a non-negative integer $n$,
if there exists a manifold $\cal M$ with boundary $\cal H$, 
a smooth function $\Omega$ defined on $\cal M$, 
and an imbedding $\Psi$ of an open subset $\h{\cal F}$ in $\h{\cal M}$
to $\cal M - \cal H$
satisfying the following conditions:

(1) $\bri:=\ppp{\cal F}\cap({\cal M}-\Psi(\h{\cal M}))$ is not
empty and $\bri\subset I^+({\cal F})$ where ${\cal F}:=\Psi({\h{\cal F}})$;

(2) $\Omega\breve{=}0$ and $\V_{a}\Omega\breve{\neq}0$, 
where $\breve{=}$ denotes the equality evaluated on $\bri$;

(3) $n^a:=\Omega^{-4}\Psi^*\h g^{ab}\V_b\Omega$ and 
    $q_{ab}:=\Omega^2(\Psi^*\h g_{ab}+\Omega^{-4}F^{-1}\V_a\Omega\V_b\Omega)$
admit smooth limits to $\bri$ with $q_{ab}$ having signature(+++) on $\bri$,
where $F:=-\Lie_n\Omega$; and

(4) $\lim_{\to\bri}\Omega^{-(2+n)}T_{\h\mu\h\nu}\brv=0$
where $\h T_{\h\mu\h\nu}:=\Psi^*[\base{\h e}{\mu}^a\base{\h e}{\nu}^b\h T_{ab}]$
in which $\{ \base{\h e}{\mu}^a \}$ and $\h T_{ab}$ are a tetrad and 
the physical energy-momentum tensor of $(\h{\cal M},\h g_{ab})$, respectively.
 
\bigskip

Henceforth, we use a tetrad consisting of 
a unit vector field $\h n^a$ that is normal to the $\Omega$-const$.$
surfaces and a triad $\{ \base{\h e}{I}^a \}_{\ssr{I=1,2,3}}$
of the metric $\h q_{ab}:=\h g_{ab}+\h n_a \h n_b$
on the $\Omega$-const$.$ surface. We denote the timelike components
with the subscript $0$ and the spacelike component 
with capital-Roman-letter subscript, e.g., $\h A^\hs0:=\h n_a \h A^a$ and
$\h A^\hs{I}:=\base{\h e}{I}_a \h A^a$.
If a tensor $A^{a\cdots b}_{\ c\cdots d}$ admits a smooth limit to
$\bri$, it is useful to define the $n$-th order term of $A^{a\cdots b}_{\ c\cdots d}$ 
as
\beq
     \up0  A^{a\cdots b}_{\ c\cdots d} := \lim_{\to\bri} A^{a\cdots b}_{\ c\cdots d}
        \hspace{4ex}\mbox{and}\hspace{4ex}
     \up n A^{a\cdots b}_{\ c\cdots d} := \lim_{\to\bri} \Omega^{-n}
            (A^{a\cdots b}_{\ c\cdots d}
               -\sum_{\ell=0}^{n-1}\up\ell A^{a\cdots b}_{\ c\cdots d}\,\Omega^\ell)
        \hspace{4ex}\mbox{for $n\geq1$}.  \label{eq:DEFexpansion}
\eeq
This definition of the $n$-th order terms of a tensor implies that in
the vicinity of $\bri$, $A^{a\cdots b}_{\ c\cdots d}$ can be expanded as 
\beq
     A^{a\cdots b}_{\ c\cdots d} 
            = \sum_{n=0}^\infty \up{n} A^{a\cdots b}_{\ c\cdots d}\,\Omega^n.
\eeq
Since all the equations appearing in the following discussion
are those on $\cal M$, 
unless it may cause ambiguity,
we omit hereafter
$\Psi^*$ in front of the tensors defined on 
$\h{\cal M}$ for brevity.

Before we examine the properties of {\sc ati-}$n$ spacetimes,
we introduce some valuable tensors.
First, the projection operator with respect to the $\Omega$-const$.$ surface
can be introduced as 
\beqn
     {q^a}_b &:=& \sum_{I=1,2,3} \base{\h e}{I}^a\base{\h e}{I}_b \nnb
              ={\delta^a}_b + F^{-1}n^a\V_b \Omega.
\eeqn
This operator admits a smooth limit to $\bri$ by virtue of the
definition of an {\sc ati-}$n$ spacetime.
Second, note that the above definition of an {\sc ati-}$n$ spacetime
implies that $\bri$ is a 3-submanifold of $\cal M$ 
with an imbedding, say $\Pi$. Hence, if a tensor field $A^{a\cdots b}_{\ c\cdots d}$
is tangential to $\bri$, 
or $A^{a\cdots b}_{\ c\cdots d}:={q^a}_e\cdots{q^b}_f\,{q^g}_c\cdots{q^h}_d 
A^{e\cdots f}_{\ g\cdots h}$,
it is useful to consider the tensor field 
$\Pi^* A^{a\cdots b}_{\ c\cdots d}$ defined on $\bri$.
Hereafter, we denote such a tensor field in boldface, i.e., 
$\bm A^{a\cdots b}_{\ c\cdots d}:=\Pi^* A^{a\cdots b}_{\ c\cdots d}$,
and say that $A^{a\cdots b}_{\ c\cdots d}$ induces $\bm A^{a\cdots b}_{\ c\cdots d}$
on $\bri$.

Now we explore the consequence of the definition of an {\sc ati-}$n$ spacetime
for $n=0$.
Solving the Einstein equation under the fall-off condition on 
the energy-momentum tensor, $\lim_{\to\bri}\Omega^{-2}\h T_{\h\mu\h\nu}=0$,
it is found that 
\beq
    \bm F \brv= 1 
      \mmbox{4ex}{and}{4ex}
    \bm q_{ab} \brv= \bm h_{ab} \label{eq:0th}
\eeq
in an {\sc ati-}$0$ spacetime,
where $h_{ab}=(d\chi)_a(d\chi)_b
+\sinh^2\chi[(d\theta)_a(d\theta)_b+\sin^2\theta(d\phi)_a(d\phi)_b]$ 
is the 3-metric of the unit timelike 3-hyperboloid. 
(For the details of the derivation, see \cite{GS}.)
Because eq.(\ref{eq:0th}) is a gravitational structure 
common to all the {\sc ati}-$0$ spacetimes,
we call it the {\em zero-th order asymptotic structure}.
Using conformal time $\eta:=\ln\Omega$, these results imply
\beq
    \h g_{ab}= \upp0\h g_{ab}+ \Omega\upp1\h g_{ab}+ O(\Omega^2)
           \mmbox{7ex}{where}{5ex}
    \upp0\h g_{ab}=  (e^{-\eta})^2 \left[
                                        - (d\eta)_a(d\eta)_b + h_{ab}
                                  \right]\label{eq:(0)gab=}
\eeq
in which $\upp n \h g_{ab}$ is defined by
\beq
      \upp n \h g_{ab} := \sum_{\mu,\nu}
           \base{\h e}{\mu}_a\base{\h e}{\nu}_b \upp n g_{\h\mu\h\nu} 
\eeq
with a function $g_{\h\mu\h\nu}:=\base{\h e}{\mu}^a\base{\h e}{\nu}^b \h g_{ab}$ 
that admits a smooth limit to $\bri$ and thus is expanded in the
manner described in eq.(\ref{eq:DEFexpansion}).
$\upp0\h g_{ab}$ is a metric of the Milne universe and is equivalent to  
the metric of a Minkowski spacetime, $\h g_{ab}^{\sf{\scriptscriptstyle{MIN}}}$. 
In other words, eq.(\ref{eq:(0)gab=}) tell us
that an {\sc ati-0} spacetime is an asymptotically Minkowski spacetime:
\beq
     \h g_{ab} = \h g_{ab}^{\sf{\scriptscriptstyle{MIN}}} +O(\Omega)
                \label{eq:min}
\eeq
Hence, it is no surprise that the Riemann tensor asymptotically vanishes
in such a spacetime.
The trace part of the Riemann tensor asymptotically vanishes by virtue of
the fall-off condition on the energy-momentum tensor.
The traceless part, or the Weyl tensor $\h C_{ambn}$,
can be best investigated by decomposing 
the tensor into the electric part 
$\h E_{ab}:=\h C_{ambn}\h n^m \h n^n$ and 
the magnetic part $\h B_{ab}:={}^* \h C_{ambn}\h n^m \h n^n$
where ${}^* \h C_{ambn}$ denotes the dual of the 2-form $\h C_{\s[am\s]bn}$.
In terms of $\h q_{ab}$ and $\h n^a$, $\h E_{ab}$ and $\h B_{ab}$ are given by
\beqn
      \hat{E}_{ab} & = & \h K_{ar}\h K^r{}_b -\Lie_{\h n}\h K_{ab} 
        + \h D_{\s(a} \h a_{b\s)} + \h a_a \h a_b
        + \frac12(\h q^r_a \h q^s_b - \h q_{ab}\h n^r \h n^s)\h L_{rs}\nnb\\
      \hat{B}_{ab} & = & \h\EEE_{ra}{}^s\h D^r\h K_{bs} 
                        + \frac12\h\EEE_{ab}{}^r\h n^s\h L_{rs}
\eeqn
where $\h K_{ab}:=\frac12\Lie_{\h n}\h q_{ab}$,
$\h L_{ab}:=\h R_{ab}-\frac16\h R\h g_{ab}$, 
$\h a_a:=\h q_{ar}\h n^s\V_s \h n^r$, and $\h D_a$ and $\h\EEE_{abc}$
are the derivative operator and the volume element associated with $\h q_{ab}$,
respectively.
Using the definitions of $n^a$ and $q_{ab}$, and imposing the fall-off
condition on the energy-momentum tensor,
it can be found that both admits a smooth limit to $\bri$
and thus can be expanded by the manner described in
eq.(\ref{eq:DEFexpansion}). Their leading terms are
\beq
    \0\bm E_{ab}\brv=0 \mmbox{7ex}{and}{5ex} \0\bm B_{ab}\brv=0.
\eeq
Simple calculation shows that the behaviors of 
$\up{n}\bm E_{ab}$ and $\up{n}\bm B_{ab}$ for
$n\geq1$ depends on that of the higher order energy-momentum tensor,
which is arbitrary in an {\sc ati}-$0$ spacetime.
(See \cite{GS} for the derivations.)

\section{the first-order asymptotic structure}\label{sec:first}

In this section, we derive the first order asymptotic structure,
that is, a structure which is possessed by all the {\sc ati-}$1$ spacetimes
but not by {\sc ati-}$0$ spacetimes.
The part of the structure, i.e., the $O(\Omega)$ term of $F$, has been already 
derived in \cite{GS}. Here, we treat the whole structure in a systematic way.
The point is that 
the first order asymptotic structure may be considered as 
the perturbation to the Milne universe, eq.(\ref{eq:(0)gab=}), 
where $\Omega$ plays the role of a small parameter of the perturbation.
Hence, we can apply the technique of the cosmological perturbation\cite{KS,MH} to
derive the first order asymptotic structure.

The energy-momentum tensor of an {\sc ati-}$1$ spacetime satisfies
a stronger fall-off condition, $\lim_{\to\bri}\Omega^{-3}\h T_{\h\mu\h\nu}=0$,
than that of an {\sc ati-}$0$ and thus the behavior of 
asymptotic gravitational fields is constrained stronger.
In other words, {\sc ati-}$1$ spacetimes possess more asymptotic
gravitational structure in common.
The structure can be derived by solving the Einstein equation under the condition
$\lim_{\to\bri}\Omega^{-3}\h T_{\h\mu\h\nu}=0$.
To obtain the equation, we first decompose the first-order metric $\upp1\h g_{ab}$ as
\begin{eqnarray}
     \upp1\h g_{ab} =  (e^{-\eta})^2
                    \left[
                            \1 F (d\eta)_a(d\eta)_b - 2\1\beta_{(a}(d\eta)_{b)}
                          -2\1\psi h_{ab} + 2\upp1\chi_{ab} 
                    \right] \label{eq:1gab}
\end{eqnarray}
where $\1\beta_a$ and $\upp1\chi_{ab}$ are tangential to the $\Omega$-const$.$
surfaces, i.e.,
$q_a{}^b\1\beta_b=\1\beta_a$ and $q_a{}^c q_b{}^d \upp1\chi_{cd}=\1\chi_{ab}$;
and $\upp1\chi_{ab}$ is traceless, i.e., $q^{ab}\upp1\chi_{ab}=0$.
With these quantities, 
the Einstein equation induces on $\bri$
the following set of differential equations in an {\sc ati-}$1$ spacetime
satisfying $\lim_{\to\bri}\Omega^{-3}\h T_{\h\mu\h\nu}=0$:
\beqn
    3\1 \bm F +2\bm\tri\1\bm\psi - 2 \bm D_m\1\bm\beta^m 
         +\bm D^m \bm D_n \1\bm\chi_m{}^n \breve= &0& \nnb\\
   \bm D_a(\1\bm F +2\1\bm\psi)+\frac12(\bm\tri-2)\1\bm\beta_a 
          - \frac12 \bm D_a (\bm D_m \1\bm\beta^m)
                      +\bm D_m \1\bm\chi_a{}^m \breve= &0& \nnb\\
    (\bm h_{ab} \bm\tri -\bm D_a\bm D_b)(\1\bm F + 2\1\bm\psi)
      + 2\bm D_{\s(a}\1\bm\beta_{b\s)} + 2 (\bm\tri+3)\1\bm\chi_{ab}
                 \qquad\qquad&&\nnb\\\qquad
              - 2 \bm h_{ab}\bm D_m\1\bm\beta^m
              - 4 \bm D_{\s(a}\bm D^m\1\bm\chi_{b\s)m}
              + 2 \bm h_{ab} \bm D^m \bm D_n \1\bm\chi_m{}^n \breve= &0&
                      \label{eq:Einstein1}
\eeqn
where $\bm D_a$ is the derivative operator associated with the metric
$\bm h_{ab}$ of $\bri$, and $\bm\tri:=\bm D^a\bm D_a$.
To simplify the equation above, we consider the Poisson gauge in which 
$\1\bm\beta_a$ is transverse and $\upp1\bm\chi_{ab}$ is transverse-traceless.
Noting that the quantities are transformed as 
\beqn
  \1 \ov{\bm F}       &\brv=& \1 \bm F \hspace{20.2ex}
  \1 \ov{\bm\beta}_a   \brv=  \1\bm\beta_a+\bm D_a \bm T-\bm L_a\nnb\\
  \1 \ov{\bm\psi}     &\brv=& \1\bm\psi + \bm T -\frac13\bm D^a \bm L_a  \hspace{5ex}
  \1 \ov{\bm\chi}_{ab} \brv=  \1\bm\chi_{ab} + \bm D_{\s(a} \bm L_{b\s)}
                                             -\frac13\bm h_{ab} \bm D_m \bm L^m
                     \label{eq:trans}
\eeqn
under a gauge transformation generated by $\xi^a=\Omega T (\ppp_\eta)^a + \Omega L^a$,
we find that the Poisson gauge can be always chosen  
if we set
\beqn
     &\bm T :\brv= -\frac12\bm\tri^{-1}
          [3(\bm\tri-3)^{-1}\bm D^m\bm D^n \upp1\bm\chi_{mn}+2\bm D^a\1\bm\beta_a]&\nnb\\
     &\bm L_a :\brv= \frac12(\bm\tri-2)^{-1} 
         \{ \bm D_a[(\bm\tri-3)^{-1}\bm D^m\bm D^n \upp1\bm\chi_{mn}]
                  -4\bm D^m \upp1\bm\chi_{ma} \},&
\eeqn
for the generator $\xi^a=\Omega T (\ppp_\eta)^a + \Omega L^a$,
which satisfy
 $\bm D^a \upp1\bm\chi_{ab}+\bm D^a\bm D_{\s(a}\bm L_{b\s)}
-\frac13\bm D_b(\bm D_m \bm L^m)\brv= 0$ and 
$\bm D^a\1\bm\beta_a + \bm\tri\bm T-\bm D^a \bm L_a\brv=0$.
Note that this Poisson gauge is preserved under the trasformation generated by
$\xi^a=\Omega T (\ppp_\eta)^a + \Omega L^a$ where $T$ and $L_a$ satisfies
\beq
     \bm\tri\bm T - \bm D^m \bm L_m \brv= 0 
               \mmbox{3ex}{and}{3ex}
     (\bm\tri-2)\bm L_b +\frac13\bm D_b\bm D_m \bm L^m \brv= 0. \label{eq:Gfree}
\eeq
In this gauge, the Einstein equation (\ref{eq:Einstein1})
simplifies to 
\begin{subeqnarray}
   &3\1{\bm F} + 2\bm\tri\1{\bm\psi}             \brv=0&  \slabel{eq:E2a}\\
   &\bm D_a(\1{\bm F} + 2\1{\bm\psi})
              +\frac12(\bm\tri-2)\1\bm\beta_a^\T \brv=0&  \slabel{eq:E2b}\\
   & (\bm h_{ab}\bm\tri-\bm D_a\bm D_b)
             (\1{\bm F}+2\1{\bm\psi})
      + 2 \bm D_{\s(a} \1{\bm\beta}_{b\s)}^\T
      + 2 (\bm\tri+3)\1{\bm\chi}_{ab}^{\T\T}     \brv=0&  \slabel{eq:E2c}
\end{subeqnarray}
where the over-bar is omitted which shows that the quantity is transformed
and the subscripts $T$ and $TT$ denote that the quantities are transverse and
transverse-traceless, respectively.

First, we derive the scalar $\1\bm F$.
Subtracting eq.(\ref{eq:E2a}) from the divergence of eq.(\ref{eq:E2b}), we obtain 
%
\begin{equation}
    (\bm\tri-3)\1\bm F \breve= 0. \label{eq:(D-3)F=0}
\end{equation}
%
The general solution of the above equation can be derived 
by first expanding the function $\1{\bm F}$ as
%
\begin{equation}
      \1 \bm F  (\bm\chi,\bm\theta,\bm\phi)\brv=
            \sum \bm a_{\ell m} \1 \bm F^{\ell m}(\bm\chi,\bm\theta,\bm\phi)
           \hspace{3ex}\mbox{where}\hspace{3ex}
        \1 \bm F^{\ell m} (\bm\chi,\bm\theta,\bm\phi):=
        \bm T^\ell_0(\bm\chi)\bm Y^{\ell m}(\bm\theta,\bm\phi)  \label{eq:F}
\end{equation}
%
in which the summation is taken over ${\ell\in\ZZ}$ and ${|m|\leq|\ell|}$,
and $\bm Y^{\ell m}(\bm\theta,\bm\phi)$ are the 2-dimensional spherical harmonics.
Substituting eq.(\ref{eq:F}) into eq.(\ref{eq:(D-3)F=0}),
it is found that $\bm T^\ell_0$ is given by
\beq
     \bm T^\ell_0(\chi) = \bm{\cal P}^\ell_2(\chi) \label{eq:To=P}
\eeq
where the functions $\bm{\cal P}^\ell_n(\chi)$ is defined by
%
\begin{equation}
  \bm{\cal P}^\ell_n(\chi) := \frac1{\sqrt{\sinh\bm\chi}} 
                      \bm P^{\ell+\frac12}_{n-\frac12}(\cosh\bm\chi), \label{eq:DefP}
\end{equation}
%
and satisfies
\beq
     {\bm{\cal P}^\ell_n}''(\bm\chi) +\frac2{\tanh\bm\chi}{\bm{\cal P}^\ell_n}'(\bm\chi) 
  - \left( n^2-1 +\frac{\ell(\ell+1)}{\sinh^2\bm\chi}\right)\bm{\cal P}^\ell_n(\bm\chi)
             \brv=0\label{eq:P}
\eeq
where the prime ${}'$ denotes the derivative with respect to $\bm\chi$;
and $P^{\mu}_{\nu}(z)$ is an associated Legendre function normalized as
%
\begin{equation}
      P^{\mu}_{\nu}(z):= 
          \frac{1}{ \Gamma(1-\mu) }\left( \frac{z+1}{z-1} \right)^{\mu/2} 
              F \left( -\nu,\nu+1,1-\mu; \frac{1-z}{2} \right) 
      \quad \mbox{for} \quad z>1.
\end{equation}
%
(Note here that unconventional choice of $\ell$ is taken.
It is related to the conventional choice $\ell_{\sf{\scriptscriptstyle{c}}}$ by
$\ell=-\ell_{\sf{\scriptscriptstyle{c}}}-1$,
and thus it is $\ell\leq-1$ terms that are regular.
The reason for the chioce is to have the coefficinent $a_{\ssr{00}}$ in
eq.(\ref{eq:F}) correspond to 
the mass of a spacetime. See eq.(\ref{eq:schw}).)

Second, we derive the scalar $\1\bm\psi$.
Adding eq.(\ref{eq:(D-3)F=0}) to eq.(\ref{eq:E2a}), we obtain
\beq
     \bm\tri(\1\bm F+2\1\bm\psi)\brv=0.
\eeq
Let $\bm A$ be the solution of $\bm\tri\bm A\brv=0$. Then, $\1\bm\psi$
may be given by
\beq
    \1\bm\psi\brv=-\frac12\1\bm F +\frac{\bm A}2.
\eeq
Now, perform the gauge trasfomation generated by $\xi^a=\Omega T(\ppp_\eta)^a$ where
$\bm T\brv=-\bm A/2$. This transformation preserves the Poisson gauge,
which can be seen from eq.(\ref{eq:Gfree}), and 
simplifies $\1\bm\psi$:
\beq
    \1{\bm\psi}\brv= -\frac12\1{\bm F}.            \label{eq:psi}
\eeq

Next, we consider the vector field $\1{\bm\beta}_a^{\T}$.
Substitution of eq.(\ref{eq:psi}) into eq.(\ref{eq:E2b}) yields
\beq
     (\bm\tri-2)\1\bm\beta_a^{\T}\brv=0.
\eeq
Hence, together with the fact $\1{\bm\beta}_a^{\T}$ is transverse, we
see from eq.(\ref{eq:Gfree}) that the gauge trasformation generated by 
$\xi^a=\Omega L^a$ where $\bm L_a\brv=\1{\bm\beta}_a^{\T}$
preserves the Poisson gauge, and results in
\beq
     \1\bm\beta_a^\T\brv=0.     \label{eq:beta=0}
\eeq

Finally, we derive the tensor field $\upp1{{\bm\chi_{ab}^{\T\T}}}$.
Substituting eqs.(\ref{eq:psi})(\ref{eq:beta=0}) into eq.(\ref{eq:E2c}), 
we obtain
\beq
     (\bm\tri+3)\upp1\bm\chi_{ab}^{\T\T}\brv=0.\label{eq:Dchi=0}
\eeq
To solve this equation, we consider the spherical harmonic expansion
again. As generally done, we decompose $\upp1\bm\chi_{ab}^{\T\T}$ 
into the even (or electric-type) parity part 
$\upp1{\bm\chi}_{ab}^{\s{TT(+)}}$
and the odd (or magnetic-type) parity part
$\upp1{\bm\chi}_{ab}^{\s{TT(-)}}$:
$\upp1{{\bm\chi_{ab}^{\T\T}}}\brv= \upp1\check{\bm\chi}_{ab}^{\s{TT(+)}}
                                     + \upp1\check{\bm\chi}_{ab}^{\s{TT(-)}}$.
In the present case, the even parity part that satisfies
eq. (\ref{eq:Dchi=0}) is found to be\cite{TS} 
\beq
     \upp1\bm\chi_{ab}^{\s{TT(+)}}\brv=(\bm D_a\bm D_b - \bm h_{ab})\bm X
\eeq
where $\bm X$ is a function that satisfies $(\bm\tri-3)\bm X\brv=0$.
On the other hand, the odd parity part is found to be\cite{To}
\beq
     \upp1{\bm\chi}_{ab}^{\s{TT(-)}}\brv=\sum_{\ell\neq0}\bm b^{\s{(-)}}_{\ell m}
                                      \,\upp1{\bm\chi}_{ab}^{\s{TT(-)}\ell m}
                                           \label{eq:chi}
\eeq
where the summation is taken over ${\ell\in\ZZ}$ for $\ell\neq0$, ${|m|\leq|\ell|}$; 
and 
\beq
   \upp1{\bm\chi}_{\chi\chi}^{\s{TT(-)}\ell m}:\!\brv=0, \hspace{3ex}
   \upp1{\bm\chi}_{\chi \s A}^{\s{TT(-)}\ell m}:\!\brv=
                          \bm T_1^\ell(\bm\chi)\bm\EEE_{\s A}{}^{\s B}
                          \bm{\cal D}_{\s B}\bm Y^{\ell m}
             \mmbox{3ex}{and}{3ex}
   \upp1{\bm\chi}_{\s{AB}}^{\s{TT(-)}\ell m}:\!\brv= 
		            \bm T_2^\ell(\bm\chi)\bm\EEE^{\s C}{}_{\s{(A}}
			    \bm{\cal D}_{\s{B)}}\bm{\cal D}_{\s C}\bm Y^{\ell m}
                    \label{eq:chi-lm}
\eeq
in which functions $\bm T_1^\ell(\bm\chi)$ and $\bm T_2^\ell(\bm\chi)$ 
are given by
\beq
     \bm T_1^\ell(\bm\chi) \brv= \bm{\cal P}^\ell_0(\bm\chi) 
             \mmbox{5ex}{and}{5ex}
     \bm T_2^\ell(\bm\chi) \brv= \frac{\sinh^2\bm\chi}{(\ell-1)(\ell+2)}
             \left[ \ppp_\chi +2 \frac{\cosh\bm\chi}{\sinh\bm\chi} \right]
                                        \bm{\cal P}^\ell_0(\bm\chi)
                 \mmbox{2ex}{for\  $\ell\neq1,-2$.}{0ex}\label{eq:T(chi)}
\eeq
(Here, there is no need to derive $\bm T_2^\ell(\bm\chi)$ 
for $\ell=1$ or $\ell=-2$ since 
$\bm\EEE^{\s C}{}_{\s{(A}}\bm{\cal D}_{\s{B)}}\bm{\cal D}_{\s C}\bm Y^{\ell m}$
vanishes for these values of $\ell$.)
Next, consider the gauge transformation generated 
by $\xi^a=\Omega T (\ppp_\eta)^a + \Omega L^a$ 
where $\bm T :\brv= -\bm X $ and $\bm L_a :\brv=-\bm D_a \bm X$.
This transformation leaves $\1\bm\beta_a$ and $\1\bm\psi$ unchanged
and kills the even parity part of $\upp1\bm\chi_{ab}^{\T\T}$:
\beq
    \upp1\bm\chi_{ab}^{\T\T}\brv=\sum_{\ell\neq0} \bm b^{\s{(-)}}_{\ell m}
                                      \,\upp1{\bm\chi}_{ab}^{\s{TT(-)}\ell m}.
\eeq

To summarize, the solutions of the Einstein equation
(\ref{eq:Einstein1}) is given by
\beqn
     & &\1\ov{\bm F}   \brv= ~~~~~\sum\bm a_{\ell m}\1\bm F^{\ell m},
                           \hspace{10ex}
        \1\ov{\bm\beta}_a\brv= 0, \nnb\\
     & &\1\ov{\bm\psi} \brv=  -\frac12\sum\bm a_{\ell m} \1\bm F^{\ell m}   
                           \hspace{4ex} \mbox{and} \hspace{4ex}
 \upp1\ov{\bm\chi}_{ab}\brv= \sum_{\ell\neq0}\bm b^{\s{(-)}}_{\ell m}
                                 \,\upp1\bm\chi_{ab}^{\s{TT(-)}\ell m} \label{eq:1st}
\eeqn
in the suitably chosen Poisson gauge.
This is the structure of the gravitaional field common 
to all the {\sc ati-}1 spacetimes, 
and thus we define {\em the first order asymptotic structure}
as follows.

\medskip

\noindent
{\bf DEFINITION:} 

$\1\h g_{ab}$ given by eq.(\ref{eq:1gab}) is called
the {\em first order asymptotic structure} of an {\sc afti}-$1$ spacetime,
where $\1\bm F$, $\1\bm\beta_a$, $\1\bm\psi$ and $\1\bm\chi_{ab}$ takes
the form eq.(\ref{eq:1st}) on $\bri$, in the Poisson gauge.

\medskip

In such a spacetime, it can be calculated that
\beq
 \1\bm E_{ab} \brv= \frac12 (\bm D_a \bm D_b \!-\! \bm h_{ab} )\1\bm F  
       \hspace{5ex} \mbox{and}  \hspace{5ex}
 \1\bm B_{ab} \brv= \frac12\,\bm \EEE_{ra}{}^s \bm D^r \upp1\bm\chi_{bs}
                                  \label{eq:1Eab1Bab}
\eeq
and that $\up{n}\bm E_{ab}$ and $\up{n}\bm B_{ab}$ for $n\geq2$
depend on the behavior of the higher order energy-momentum tensor,
which is arbitrary in an {\sc ati-}$1$ spacetime.
(See \cite{GS} for the details of the calculation.)
Eq.(\ref{eq:1Eab1Bab}) clearly shows that $\1\bm F$ of the first order
asymptotic structure forms the first order term of the electric part of 
the Weyl tensor  and
that $\upp1\bm\chi_{ab}$ forms the magnetic part. 
In other words, if we specify the sets of coefficients $\{\bm a_{\ell m} \}$
and $\{\bm b_{\ell m} \}$, the first order terms of the electric part 
and the magnetic part are determined, respectively.

\section{asymptotic stationarity}\label{sec:stationary}

In this section, we introduce the notion of 
{\em asymptotic stationarity} of {\sc ati-}$n$ spacetimes, 
and prove that an {\sc ati-}$1$ spacetime that is asymptotically stationary
to order 2 must be an asymptotically Schwarzschild spacetime.
The plan of the proof is: 1) we derive, in the Poisson gauge, 
the reduced first asymptotic structure of an {\sc ati-}$1$ spacetime 
that is asymptotically stationary to order 2 in the lemma; 
2) we perform a suitable gauge transformation
as to show explicitly that such an asymptotic structure 
approaches asymptotically the Schwarzschild metric in the theorem.

A Killing vector field $\h\xi^a$ is a vector field with respect to which 
the Lie derivative of the metric vanishes, $\Lie_{\h\xi} \h g_{ab}=0$.
This fact motivates us to define an asymptotic Killing vector field 
and its order as follows.

\medskip

\noindent
{\bf DEFINITION:} \enskip An {\sc ati-}$m$ spacetime ($\hat{\cal M}$,$\hat{g}_{ab}$) 
is said to admit an 
{\em asymptotic Killing field} $\h\xi^a$ to order $n$ if 
%
\beq
   \lim_{\to\bri}\Omega^{-n}(\Lie_{\h\xi}\h g)_{\h\mu\h\nu} = 0\label{eq:DEFasymK}
\eeq
%
where $(\Lie_{\h\xi}\h g)_{\h\mu\h\nu}:=
\Psi^*(\base{\h e}{\mu}^a\base{\h e}{\nu}^b\Lie_{\h\xi}\h g_{ab})$. 

\medskip


Next, we consider how to define asymptotic stationarity,
using this notion of an asymptotic Killing vector field.
A vector field $\h\xi^a$ is said to be a stationary Killing vector field 
in an asymptotically flat spacetime, if $\h\xi^a$ is a timelike 
Killing field and satisfies $\h g_{ab}\h\xi^a\h\xi^b=-1$ at infinity.
Hence, we define its asymptotic correspondace as follows.
\medskip

\noindent
{\bf DEFINITION:} \enskip A vector field $\xi^a$ is said to be 
an asymptotic stationary Killing vector field to order $n$ 
of an {\sc ati-}$m$ spacetime
if $\h\xi^a$ is admitted as an asymptotic Killing vector field to order $n$
in the {\sc ati-}$m$ spacetime and satisfies
\beq
    \h g_{ab}\h\xi^a\h\xi^b \brv= -1 \label{eq:DEFtransK}
\eeq
on $\bri$.

\medskip


It is important to note that the definition of 
the asymptotic stationary Killing vector field
implies that the leading term of 
$\h\xi^{\h\mu}=\h\xi^a\base{\h e}{\mu}_a$ is of the order $\Omega^0$. 
Hence, 
$\xi^\h\mu:=\h\xi^\h\mu$ admits a smooth limit to $\bri$ and can be expanded as
\beq
    \xi^\h\mu = \sum_{n=1}^\infty \up{n}\xi^\h\mu\,\Omega^n. 
                          \label{eq:XIexpanded}
\eeq
(Here, $\up{n}\xi^\hs0$ and $\up{n}\xi^\hs{\ssr I}$ correspond to
$\up{n+2}\xi^\hs0$ and $\up{n+1}\xi^\hs{\ssr I}$ in \cite{GS}.
We do not use the notation of \cite{GS} for $\h\xi^a$,
because the above notation reflects the nature of the completion more intrinsically.)
Simple calculation shows that the function $(\Lie_\h\xi \h g)_{\hs\mu\hs\nu}$
admits a smooth limit to $\bri$, and thus can be expanded in the manner
described by eq.(\ref{eq:DEFexpansion}).

Now we are ready to prove the following lemma.
\bigskip

\noindent
{\bf LEMMA:} \enskip 
An {\sc ati-1} spacetime is asymptotically stationary to order 2
if and only if 
$\bm a_{\ell m}\brv=0$ for $\ell\neq0$ and $\bm b_{\ell m}\brv=0$.

\medskip
{\em Proof of only if:}
If an {\sc ati-1} spacetime is asymptotically stationary to order 2,
\beqn
     &\up{n}(\Lie_\h\xi \h g)_{\hs\mu\hs\nu}\brv=0 
         \hspace{2ex}\mbox{for $n\leq 2$},&      \label{eq:condOnLg}\\
     &\h g_{ab} \h\xi^a \h\xi^b \brv= -1&         \label{eq:condOnXi}
\eeqn
hold. First, we note that eq.(\ref{eq:condOnXi}) is equivalent to 
\beq
     -(\0\bm\xi^\hs0)^2 + \0\bm \xi^a \0\bm \xi_a\brv=-1 \label{eq:0xiB}
\eeq
where $\up\ell\bm\xi^a:=\up\ell\bm\xi^\hs{K}\base{\bm e}{K}^a$ and
$\{\base{\bm e}{I}_a\}_{\ssr{I=1,2,3}}$ is a triad of $\bm h_{ab}$.
Second, we simplify the $n\leq1$ part of eq.(\ref{eq:condOnLg}).
In an {\sc ati-}1 spacetime, 
the $O(\Omega^0)$ and $O(\Omega^1)$ terms of $(\Lie_\h\xi \h g)_{\hs\mu\hs\nu}$
are given by
\beqn
 &\0(\Lie_\h\xi \h g)_{\hs0\hs0}    \brv=0,&  \hspace{14ex}
  \0(\Lie_\h\xi \h g)_{\hs0\hs{I}}  \brv=0,   \hspace{26ex}
  \0(\Lie_\h\xi \h g)_{\hs{I}\hs{J}}\brv=0,   \nnb\\
 &\1(\Lie_\h\xi \h g)_{\hs0\hs0}    \brv=0,&  \hspace{4ex}
  \1(\Lie_\h\xi \h g)_{\hs0\hs{I}}  \brv= 
                \left[\0\bm\xi_a-\bm D_a \0\bm\xi^\hs0\right]\base{\bm e}{I}^a,  
                                         \hspace{3ex}\mbox{and}\hspace{2ex}
  \1(\Lie_\h\xi \h g)_{\hs{I}\hs{J}}\brv= \left[\bm D_{\s(a}\!\!\0\bm \xi_{b\s)} 
                                            - \!\!\0\bm\xi^\hs0\bm h_{ab}\right]
                                         \base{\bm e}{I}^a\base{\bm e}{J}^b.   \nnb
\eeqn
Hence, the $n\leq1$ part of eq.(\ref{eq:condOnLg}) is equivalent to
\beq
    \0\bm\xi_a-\bm D_a \0\bm\xi^\hs0\brv=0
        \mmbox{3ex}{and}{3ex}
    \bm D_{\s(a}\!\!\0\bm \xi_{b\s)} - \!\!\0\bm\xi^\hs0\bm h_{ab}\brv=0.\label{eq:0xiA}
\eeq
Solving eqs.(\ref{eq:0xiB})(\ref{eq:0xiA}) simultaneously, we find that 
\beq
    \0\bm\xi^\hs0 \brv= \cosh\bm\chi \hspace{4ex}\mbox{and}\hspace{4ex}
    \0\bm\xi^\hs{K}\base{\bm e}{K}^a \brv=\sinh\bm\chi(\ppp_\chi)^a.  \label{eq:choice}
\eeq
Next, we consider the $n=2$ part of eq.(\ref{eq:condOnLg}).
Let us choose the Poisson gauge in which the first order asymptotic
structure takes the form eq.(\ref{eq:1st}) in an {\sc ati-1} spacetime.
Because
\beqn
&\up2(\Lie_\h\xi \h g)_{\hs0\hs0}     \brv=& 
                                - \1\bm\xi^\hs0 
				+ \frac12\left( \0\bm\xi^\hs0 + \0\bm\xi^m\bm D_m
                                        \right)\1\bm F
                                         \hspace{4ex}\mbox{and} \nnb\\
&\up2(\Lie_\h\xi \h g)_{\hs0\hs{I}}   \brv=& \left[
                             2 \1\bm\xi_a -\bm D_a \1\bm\xi^\hs0 
                             + \left(\bm D_a \0\bm\xi^\hs0-\0\bm\xi_a\right)\1\bm F
                             -2\0\bm\xi_m\upp1\bm\chi_a\!{}^m
                                            \right]\base{\bm e}{I}^a, 
\eeqn
the $n=2$ part of eq.(\ref{eq:condOnLg}) for $\h\mu=\h 0$ is equivalent to
\beqn
    \1\bm\xi^\hs0 \brv= \frac12\left( \0\bm\xi^\hs0 + \0\bm\xi^m\bm D_m
                                        \right)\1\bm F
                               \hspace{3ex}\mbox{and}\hspace{3ex}
    \1\bm\xi^\hs{K} \base{e}{K}^a \brv=  \frac12\left[\bm D_a \1\bm\xi^\hs0 
                             - \left(\bm D_a \0\bm\xi^\hs0-\0\bm\xi_a\right)\1\bm F
                             +2\0\bm\xi_m\upp1\bm\chi_a\!{}^m \right].
\label{eq:choice2}
\eeqn
Using eqs.(\ref{eq:choice})(\ref{eq:choice2}),
we find that $\up2(\Lie_\h\xi \h g)_{\hs{I}\hs{J}}$ are given by 
\beq
     \up2(\Lie_\h\xi \h g)_{\hs{I}\hs{J}}\brv= 2\sinh\bm\chi
         \left[  \straightup1\bm{\cal L}_{ab} \1\bm F 
               + \straightup2\bm{\cal L}_{\s(a}\!{}^m \,\upp1\bm\chi_{b\s)m}
        \right] \base{\bm e}{I}^a\base{\bm e}{J}^b \label{eq:2Lij}
\eeq
where the derivative operators $\straightup1\bm{\cal L}_{ab}$ and $\straightup2\bm{\cal L}_{ab}$ 
are given by
\beq
 \straightup1\bm{\cal L}_{ab}
     :\brv=  \frac3{\tanh\bm\chi}(\bm D_a\bm D_b-\bm h_{ab})
             + (\ppp_\chi)^m\bm D_{\s(a}\bm D_{b\s)}\bm D_m
             - (d\bm\chi)_{\s(a}\bm D_{b\s)} \hspace{2ex} \mbox{and} \hspace{2ex}
 \straightup2\bm{\cal L}_{ab} :\brv= 2(d\bm\chi)_p
               \bm h_a\!{}^{\s[p}\bm h_b\!{}^{m\s]}\bm D_m,
\eeq
respectively. 
By the definition of $\1\bm F$ and $\1\bm\chi_{ab}$,
$\up2(\Lie_\h\xi \h g)_{\hs{I}\hs{J}}$ vanishes if and only if
$\straightup1\bm{\cal L}_{ab}\1\bm F$ and 
$\straightup2\bm{\cal L}_{\s(a}\!{}^s \,\upp1\bm\chi_{b\s)s}$ vanish independently.
With the help of eqs.(\ref{eq:F})(\ref{eq:1st}), 
$\straightup1\bm{\cal L}_{ab}\1\bm F\brv=0$ can be rewritten as
\beqn
   \sum \bm a_{\ell m}
& & \left[\hspace{1ex}
       \left( {\bm T^\ell_0}''' +\frac{3{\bm T^\ell_0}''}{\tanh\bm\chi}
                -{\bm T^\ell_0}'-\frac{3\bm T^\ell_0}{\tanh\bm\chi}
      \right) (d\bm\chi)_a(d\bm\chi)_b 
   + 2 \left( {\bm T^\ell_0}''-\frac{{\bm T^\ell_0}'}{\tanh\bm\chi}
                -(5+\frac4{\sinh^2\bm\chi})\bm T^\ell_0
      \right) (d\bm\chi)_{\s(a}\bm{\cal D}_{b\s)}
          \right. \nnb\\
& & \hspace{2ex}
   +\sinh\bm\chi\cosh\bm\chi
       \left( {\bm T^\ell_0}''+\frac{2{\bm T^\ell_0}'}{\tanh\bm\chi}-3\bm T^\ell_0
                  -\frac{\ell(\ell+1)}{2\sinh\bm\chi\cosh\bm\chi}
                          ({\bm T^\ell_0}'+\frac{\bm T^\ell_0}{\tanh\bm\chi})
      \right) (d\bm\sigma)_{ab}
                  \nnb\\
& & \hspace{2ex}\left.
   + \left({\bm T^\ell_0}'+\frac{\bm T^\ell_0}{\tanh\bm\chi}\right)
        (\bm{\cal D}_a\bm{\cal D}_b-\frac12(d\bm\sigma)_{ab}\bm{\cal D}^c\bm{\cal D}_c )
          \right]\bm Y^{\ell m} \brv= 0.   \label{eq:LijScalar}
\eeqn
We first consider the $\ell\neq0$ terms of the above equation.
Noting that all the components are independent and
using eqs.(\ref{eq:To=P})(\ref{eq:P}), it is found that 
{\em if} $\bm a_{\ell m}\brv{\neq}0$
the above equation (\ref{eq:LijScalar}) is equivalent to
\beqn
  \frac{\ell(\ell+1)}{\sinh^2\bm\chi}
             ({\bm{\cal P}^\ell_2}'+\frac{\bm{\cal P}^\ell_2}{\tanh\bm\chi}) \brv= 0,
          \hspace{2ex}
  \left( \frac{\ell^2+\ell-4}{\sinh^2\bm\chi}-2 \right)\bm{\cal P}^\ell_2
             -3{\bm{\cal P}^\ell_2}'\brv=0, 
          \hspace{2ex}
  \frac{\ell(\ell+1)}{\sinh^2\bm\chi}\bm{\cal P}^\ell_2 \brv= 0 
          \hspace{2ex}\mbox{and}\hspace{2.5ex}
    {\bm{\cal P}^\ell_2}'+\frac{\bm{\cal P}^\ell_2}{\tanh\bm\chi} \brv= 0.
               \label{eq:LijScalarReduced}
\eeqn
Apparently, there is no integer $\ell$ that is not equal to $0$ and
that satisfies eqs.(\ref{eq:LijScalarReduced}), simultaneously. 
Therefore, $\bm a_{\ell m}\brv=0$ for $\ell\neq0$.
Noting that $\bm{\cal D}_a\bm Y^{\ell m}$ vanishes for $\ell=0$,
we find that all the components of the $\ell=0$ terms of the right-hand
side of eq.(\ref{eq:LijScalar}) vanishes. 
Therefore, $\bm a_{\ssr{00}}$ can take arbitrary value.
Next consider $\straightup2\bm{\cal L}_{\s(a}\!{}^s \,\upp1\bm\chi_{b\s)s}\brv=0$.
This equation is satisfied if and only if 
\beq
     \sum_{\ell\neq0} \bm b_{\ell m} \left[ 
        ( {\bm T^\ell_1}'+\frac{\bm T^\ell_1}{\tanh\bm\chi})
                         (d\bm\chi)_{\s(a}\bm\EEE_{b\s)}{}^r\bm{\cal D}_r 
      + ( {\bm T^\ell_2}'-\frac{\bm T^\ell_2}{\tanh\bm\chi} -\bm T^\ell_1 )
                        \bm\EEE^r{}_{\s(a}\bm{\cal D}_{b\s)}\bm{\cal D}_r 
                          \right]\bm Y^{\ell m} \brv= 0. \label{eq:LijTensor}
\eeq
With the same reasoning, we find that if $\bm b_{\ell m}\brv=0$ 
the above equation (\ref{eq:LijTensor}) is equivalent to
\beq
      \frac{\ell(\ell+1)}{\sinh^2\bm\chi}\bm{\cal P}^\ell_0 \brv= 0 
             \hspace{4ex}\mbox{and}\hspace{4ex}
      \frac{(\ell+2)(\ell-1)}{\sinh^2\bm\chi}\bm{\cal P}^\ell_0 \brv= 0 
             \label{eq:LijTensorReduced}
\eeq
and that there is no $\ell$ that satisfies the above equations simulataneously.
Hence, we conclude $\bm b_{\ell m}\brv=0$.$\Box$

\medskip
{\em Proof of if:}
If $\bm a_{\ell m}\brv=0$ for $\ell\neq0$ and $\bm b_{\ell m}=0$
in an {\sc ati-$1$} spacetime,
the vector field $\h\xi^a$ whose $O(\Omega^0)$ and $O(\Omega^1)$
terms are given by eqs.(\ref{eq:choice})(\ref{eq:choice2}) satisfy
$\up{n}(\Lie_\h\xi \h g)_{\hs\mu\hs\nu}\brv=0$ for $n\leq2$
and $\h g_{ab} \h\xi^a \h\xi^b \brv= -1$.
Hence,  $\h\xi^a$ is an asymptotically stationary Killing vector field
to order 2.$\Box$

\medskip

The fact that $\bm a_{\ell m}\brv=0$ for $\ell\neq0$  and
$\bm b_{\ell m}\brv=0$ means that in such an {\sc ati-$1$} spacetime,
the first asymptotic structure takes the simple form
\beq
     \1\bm F \brv= \bm a_{\ssr{00}}\1\bm F^{\ssr{00}},\hspace{4ex}
     \1\bm\psi\brv= -\frac12\bm a_{\ssr{00}}\1\bm F^{\ssr{00}},\hspace{4ex}
     \1\bm\beta\brv=0 \hspace{5ex}\mbox{and}\hspace{6ex}
     \1\bm\chi_{ab}\brv=0.\label{eq:1stASstnry}
\eeq
Before we show that such an {\sc ati-$1$} spacetime is an asymptotically
Schwarzschild spacetime, we remark an important fact relating to 
the definition of the angular-momentum of an asymptotically flat
spacetime.
To define angular-momentum, one must impose the condition that
the $O(\Omega^1)$ term of the magnetic part of the Weyl tensor
vanish\cite{AR,GS,AH}.
However, the physical meaning of the condition was left
unclear. The lemma and eqs.(\ref{eq:1Eab1Bab})(\ref{eq:1stASstnry}) tells us that 
the meaning is that the angular-momentum of an asymptotically spacetime
can be defined if the spacetime is asymptotically stationary to order 2.

\bigskip
\noindent
{\bf Theorem:} \enskip 
An {\sc ati-1} spacetime which is asymptotically stationary to order 2
is an asymptotically Schwarzschild spacetime 
with mass $a_{\ssr{00}}$ in the
sense that the metric takes the form
\beq
   \h g_{ab} = \h g_{ab}^{\sf{\scriptscriptstyle{SCH}}}+ O(r^{-2})+O(t^{-2})
           \hspace{4ex}\mbox{where}\hspace{4ex}
   \h g_{ab}^{\sf{\scriptscriptstyle{SCH}}} := -(1-\frac{2a_{\ssr{00}}}r)(dt)_a(dt)_b
                           +(1-\frac{2a_{\ssr{00}}}r)^{-1}(dr)_a(dr)_b 
                           + r^2(d\sigma)_{ab}
                   \label{eq:schw}
\eeq
in which  $t>r$.

\medskip
{\em Proof:}
From the lemma, 
in an {\sc ati-1} spacetime which is asymptotically stationary to order 2,
$\bm a_{\ell m}=0$ for $\ell\neq0$ and $\bm b_{\ell m}=0$ hold.
Thus, the first order term of the metric $\1\h g_{ab}$ takes the form
\beq
    \1\h g_{ab} = (e^{-\eta})^2 \left[\,
                    a_{\ssr{00}}\1 F^{\ssr{00}} (d\eta)_a(d\eta)_a 
                  + a_{\ssr{00}}\1 F^{\ssr{00}} h_{ab}\right].
\eeq
Under a gauge transformation generated by 
$\xi^a=\Omega T(\ppp_\eta)^a+\Omega D^a L$ where
\beqn
    \bm T :\brv= \bm a_{\ssr{00}}
                 \left(2\bm\chi\cosh\bm\chi+\sinh\bm\chi\right)
      \hspace{4ex}\mbox{and}\hspace{4ex}
    \bm L :\brv= - 4\bm a_{\ssr{00}}\sinh\bm\chi +\bm T,
\eeqn
$\1\h g_{ab}$ transforms as
\beq
  \1\h g_{ab} \mapsto (e^{-\eta})^2 \left[\,
                          a_{\ssr{00}}\1 F^{\ssr{00}} (d\eta)_a(d\eta)_a
                      - 8 a_{\ssr{00}}\cosh\chi (d\eta)_{\s(a}(d\chi)_{b\s)} 
                      + a_{\ssr{00}}\1 F^{\ssr{00}}(d\chi)_a(d\chi)_b\right].
\eeq
Then, the change of variables, $t=\Omega^{-1}\cosh\chi$ and 
$r=\Omega^{-1}\sinh\chi$, leads us to 
\beq
    O(\Omega^2)=\sinh^{-2}\chi O(\Omega^2) + \cosh^{-2}\chi O(\Omega^2)
               =O(r^{-2})+O(t^{-2}),  \hspace{4ex} \frac{r}{t}=\tanh\chi<1.
\eeq
and 
\beq
  \0\h g_{ab}+\Omega\1\h g_{ab} = \h g_{ab}^{\sf{\scriptscriptstyle{SCH}}}
                                     +O(\Omega^2).
\eeq
Hence, eq.(\ref{eq:schw}) holds.$\Box$
\bigskip


It is important to note here that asymptotically Schwarzschild
spacetimes, that are defined in eq.(\ref{eq:schw}), 
comprise the Kerr spacetime also.
This can be understood by writing the Kerr metric 
$\h g_{ab}^{\sf{\scriptscriptstyle{KER}}}$
with the coordinates $(\Omega,\chi)$ where $t=\cosh\chi$ and $r=\sinh\chi$:
\beq
          \h g_{ab}^{\sf{\scriptscriptstyle{KER}}}=
                    \h g_{ab}^{\sf{\scriptscriptstyle{SCH}}}+\Omega^2\,\upp2\h g_{ab}
                                              +O(\Omega^3) \label{eq:kerr}
\eeq
where
\beqn
& &     \upp2\h g_{ab} =  
                (4m^2-a^2\sin^2\theta)
                   \left( (d\eta)_{\s(a}(d\eta)_{b\s)}
                          -2\frac{(d\chi)_{\s(a}(d\eta)_{b\s)}}{\tanh\chi} \right)
                +4ma\frac{\sin^2\theta}{\tanh\chi}(d\phi)_{\s(a}(d\eta)_{b\s)}\nnb\\
& &\hspace{8ex}+\frac{4m^2-a^2\sin^2\theta}{\tanh\chi} (d\chi)_a(d\chi)_b 
                -4am\sin^2\theta^2(d\phi)_{\s(a}(d\chi)_{b\s)}
                +a^2\cos^2\theta(d\theta)_a(d\theta)_b
                +a^2\sin^2\theta(d\phi)_a(d\phi)_b.    \label{eq:2gKER}
\eeqn
In other words, the Kerr spacetime is a special spacetime of asymptotically
Schwarzschild spacetimes, which possesses a particular second-order
asymptotic structure, i.e.,  $\upp2\h g_{ab}$ given by eq.(\ref{eq:2gKER}).
  
\section{summary and remarks}

In this paper, we have proved that an asymptotically flat spacetime
as defined in definition~1 of sec.\ref{sec:pre} is an asymptotically
Schwarzschild spacetime in the sense of eq.(\ref{eq:schw}),
if the energy-momentum of the spacetime falls off at the rate
faster than $O(\Omega^3)$ and the spacetime is asymptotically stationary
to order 2 in the sense that $(\Lie_{\h\xi}\h g)_{\h\mu\h\nu}$ falls off 
at the rate faster than $O(\Omega^2)$ for the asymptotically timelike
vector $\h\xi^a$, $\h g_{ab}\h\xi^a\h\xi^b\brv=-1$.

Finally, we give a remark.
Although we have solved the Einstein equation (\ref{eq:Einstein1})
and obtained the first order asymptotic structure,
we did not impose any physically-suitable boundary conditions on the solutions.
Hence, the obtained first order asymptotic structure may include those that are
{\em unphysical}. For example, a solution that describes an incoming
gravitational wave from future null infinity or an outcoming wave from
the event horizon. 
In other words, {\em physically} acceptable gravitational fields
around a black hole may be obtained
only after a suitable boundary condition are imposed on the solutions.
The derivation of the physical first order asymptotic structure 
may be profitable because it may be possible to show that an {\sc ati-}1
spacetime with such a physical structure is intrinsically 
asymptotically stationary to order 1, and thus is an asymptotically
Schwarzschild spacetime.
This result is anticipated because we expet that a spacetime with
a black hole that becomes vacuum also becomes stationary,
due to the nature of the black hole.
This is an important point that should be clarified.

\vskip1cm
\centerline{\bf ACKNOWLEDGEMENT} 
We would like to thank K. Sato and Y. Suto for their 
encouragements. The discussion with T.T. Nakamura was very fruitful.
UG thanks M. Sasaki, M. Shibata, T. Tanaka and Y.Mino 
for the valuable discussions.
TS thanks Gary Gibbons and the Relativity Group at Cambridge for their
hospitality.
Finally,
we wish to thank Misao Sasaki especially for his comments on the draft of this paper.

This work is partially supported by the JSPS fellowship.


\end{document}